\documentclass[sigconf]{acmart}
\usepackage{balance}
\definecolor{mypurple}{RGB}{119, 69, 198}
\definecolor{newc}{RGB}{70, 180, 80}

\newcommand{\techreport}[1]{#1}
\newcommand{\longtechreport}[1]{}
\newcommand{\papertext}[1]{}

\usepackage{multirow}
\usepackage[vlined, ruled, linesnumbered]{algorithm2e}
\usepackage{soul}
\usepackage{tikz}
\usepackage{subcaption}
\usetikzlibrary{graphs,shapes,positioning,arrows.meta}
\usepackage{xspace}
\usepackage{url}

\usepackage{booktabs}
\usepackage{listings}
\usepackage{balance}

\setcopyright{none} 
\renewcommand\footnotetextcopyrightpermission[1]{}

\newenvironment{tightitemize}{\begin{list}{$\bullet$}{\setlength{\rightmargin}{0em}\setlength{\leftmargin}{0.5em}\setlength{\topsep}{0.5mm}\setlength{\itemsep}{0.6mm}\setlength{\itemindent}{1em}}}{\end{list}}
    
\newcommand{\stitle}[1]{\vspace{0.2em}\noindent\textbf{#1.}}

\newcommand{\takeaway}[1]{\underline{\em Takeaway:} {\em #1}}

\papertext{}

\begin{document}
\pagestyle{plain}

\title{Revisiting Prompt Engineering via Declarative Crowdsourcing}

\author{Aditya G. Parameswaran, Shreya Shankar, Parth Asawa, Naman Jain, Yujie Wang}
\affiliation{UC Berkeley \\ 
\{\url{adityagp, shreyashankar, pgasawa, naman_jain, yujie_wang}\} \url{@ berkeley.edu}}

\begin{abstract}
Large language models (LLMs)
are incredibly powerful
at comprehending and generating data in the
form of text,
but are brittle and
error-prone.
There has been an advent of toolkits
and recipes centered
around so-called prompt engineering---the
process of asking an LLM to do 
something via a series of prompts.
However, for LLM-powered 
data processing workflows, in particular,
optimizing 
for quality, while keeping cost bounded,
is a tedious, manual process.
We put forth a research agenda around {\em declarative
prompt engineering}.
We view LLMs like crowd workers
and leverage ideas from the declarative
crowdsourcing literature---including 
multiple prompting strategies, ensuring
internal consistency, 
and exploring hybrid-LLM-non-LLM 
approaches---to make prompt engineering a more principled process.
Preliminary case studies on sorting, entity resolution,
and missing value imputation demonstrate
the promise of our approach.
\end{abstract}
\maketitle

\begingroup
\renewcommand\thefootnote{}\footnote{}

\section{Introduction}

Large Language Models (LLMs), such as GPT-4, ChatGPT, Claude, 
and Bard, have taken the world by storm. 
At least part of the excitement surrounding LLMs 
has been that they show promise in new tasks for 
which they have not been
explicitly
trained for.
Now, users can simply specify instructions for new tasks
as a text {\em prompt}, and the LLM can perform
these tasks as instructed, to varying degrees of 
accuracy~\cite{liupretrain2023,weng2023prompt}. 
Developers spend long periods of 
time iterating on prompts and sequences thereof, i.e., 
performing {\em 
prompt engineering}.
Prompt engineering,
despite its name, 
is very much an art---supplemented with the rise of 
cookbooks, recipes, guides, and more---all 
offering varying degrees of help for 
how to best involve LLMs in various 
types of workflows~\cite{wsj-chatgpt,forbes-prompt-engineer,Saravia_Prompt_Engineering_Guide_2022,ding2021openprompt,bach2022promptsource, openai-cookbook}.
Specifically, for prompts that produce promising demos
on a small scale,
reliably translating such demos 
to production---as part 
of complex workflows
operating on more data 
to consistently 
accomplish a global objective---is often 
laborious and frustrating.
{\em We therefore need a set of principles
around prompt engineering 
that make robust production 
deployments possible.}

To develop principles for leveraging LLMs 
in complex workflows, we turn to the domain of crowdsourcing.
Fundamentally, {\em one can view LLMs as
noisy human oracles.}
Like humans, LLMs make mistakes, are biased and inconsistent, 
fail to precisely follow instructions, 
make up information, and answer confidently even 
if they don't know the right answer.
Thankfully, there is a rich body of literature 
spanning over a decade on
how to best leverage the crowd for various
data processing workflows, and specifically,
dealing with the fact that humans
are error-prone when answering questions. 
In particular, the database community
focused on systems and toolkits for
{\em declarative crowdsourcing}~\cite{crowddb,qurkCIDR,deco}, 
with the vision that the user simply specifies 
the high-level data processing objectives,
and the system will decompose it to the right
sequence of human tasks 
automatically, taking into account
cost, accuracy, and latency~\cite{li2016crowdsourced,marcus2015crowdsourced}.

Building on this literature, 
we propose a new research agenda around 
{\em declarative
prompt engineering}. 
We similarly envision users of LLMs
specifying their data processing objectives
at a high level with the system
decomposing it into unit calls
to one or more types of LLMs
with appropriate prompts,
acting on one or more units of data, 
orchestrating
the calls, and issuing more as needed,
until completion to the desired
degree of accuracy, 
within the specified monetary budget. 
Our focus
is on data processing to keep the scope finite,
but our techniques are broadly applicable.

Like the declarative crowdsourcing
literature, our focus is not
so much on the specific textual 
instructions for a given task---in LLM parlance, 
an individual prompt---but much more
on the underlying data processing operation.
Indeed, we are still in the early 
days of understanding how the 
textual wording of the prompt impacts accuracy---for example,
even small variations can yield 
different results~\cite{zhao2021calibrate}---or 
when and how LLMs hallucinate~\cite{holtzman2019curious}.
We instead focus on the underlying data 
processing
operation---i.e., the input
data items embedded
into the prompt, and the output data items
extracted from the text generated by the LLM,
and the relationships thereof---and consider 
how one or more such operations can be 
sequenced
as part of a broader workflow.
We take as a given textual prompts that work
well on the small scale---possibly 
drawn from one of the prompt repositories~\cite{ding2021openprompt,bach2022promptsource} or guides~\cite{Saravia_Prompt_Engineering_Guide_2022,openai-cookbook} 
and focus more
on scaling up their use for multi-step production deployments.

\vspace{-5pt}
\begin{example}[Entity Resolution]
    To illustrate the challenges with leveraging LLMs for
data processing, and how
ideas from crowdsourcing may help, 
let's consider a simple example.
Suppose we have a number of product
listings, each as individual records,
and wanted to perform entity resolution
across this set of records.
The standard approach would be 
to simply provide the entire list of
records, prefixed
with an appropriate prompt,
to the LLM,
and ask it to group these into 
distinct groups of records.
Unfortunately, LLMs have limited
context lengths (i.e., a limit on the
number of tokens they can accept as input), 
so it may be hard to fit
the entire list of records
into a prompt.
And even if one could, LLMs
often make mistakes and
hallucinate, even on tasks
with as few as 20 records,
as we will see in the following.
It's not clear how one can easily
detect, let alone 
deal with, mistakes and hallucinations.
Drawing on the crowdsourcing literature,
one could decompose this problem
into smaller tasks for the LLM, such
as asking the LLM to compare pairs
of records to see if they
are duplicates~\cite{markus-sorts-joins}.
However, this approach could be expensive,
requiring $O(n^2)$ comparisons.
An intermediary option could
be to provide the LLM
smaller groups of records to resolve
into duplicates,
but we would need to sequence the tasks
or prompts in the right way
to ensure that every record
is compared to every other record~\cite{crowder}. 
There are also other mechanisms from
the crowdsourcing literature that could help reduce cost.
We could, for example, leverage
transitivity 
to automatically determine
that two records $a$ and $b$ are
duplicates if they are, for example,
deemed to be identical to another record~\cite{wang2013leveraging}.
We could also combine LLM-based approaches
with other non-LLM-based proxies, 
such as a cheaper embedding-based model,
that helps identify potential
duplicates or non-duplicates---and
only use the LLM for the ``confusing'' cases.
We could even ask the LLM to synthesize
blocking rules~\cite{Gokhale:2014wv},
or derive features to help build a cheaper 
proxy model~\cite{markus-sorts-joins}.
Overall, it's clear that there
are many ideas from the declarative
crowdsourcing literature
that can be leveraged to provide 
valuable points along the cost-accuracy tradeoff curve
for this problem.
\end{example}
\vspace{-5pt}

\techreport{In this work, we focus on closed-ended tasks,
such as sorting, filtering, or categorizing
a set of data items,
rather than open-ended ones. 
There is indeed work on open-ended crowdsourcing
that may be applicable; e.g., Turkomatic~\cite{Kulkarni:2012gi}
uses the crowd to break down an open-ended task into sub-tasks that
are then each individually crowdsourced. 
However, the emergent properties
of LLMs for open-ended tasks are still 
largely being discovered~\cite{bubeck2023sparks}
and so we plan to investigate that as future work. 
Finally, we note that we do not believe
that LLMs are a replacement for human
expertise and domain knowledge; however,
there are finite, closed-ended tasks that we 
believe LLMs can offer a compelling
low-cost approach. }

\stitle{Related Work}
To the best of our knowledge,
{\em ours is the first paper to view LLMs
as noisy human oracles,
and leverage principles from declarative crowdsourcing
to craft better data processing workflows
powered by LLMs}.
Over the past six months, 
there has been an advent of software solutions
for LLM-based data processing applications, 
recognizing that LLMs are error-prone and that LLM-based workflows often
involve multiple LLM invocations.
These solutions, including LangChain (python.langchain.com), 
Guardrails (getguardrails.ai),  LlamaIndex (www.llamaindex.ai),
and Prompt\-Chainer~\cite{wu2022promptchainer}, however,
are wrappers around LLM APIs
that enable users to implement multi-step LLM workflows, combine with external knowledge bases, and check syntactic correctness of outputs,
but offer no guidance on how
to craft LLM-based data processing 
workflows to meet certain desired objectives. \longtechreport{They
therefore fall
well short of full declarativity.
For example:
\begin{tightitemize}
\item LangChain (\url{https://python.langchain.com/})
enables users to manually compose multi-step LLM pipelines,
as  {\em components}, with predefined
prompt templates, but doesn't provide any support for
optimizing for overall accuracy or cost, 
apart from retries
if individual outputs don't meet some constraints. 
A recent extension helps record histories of LangChain invocations but offers
little in way of optimization.
\item Guardrails (\url{https://shreyar.github.io/guardrails/})
asks users to provide a schema
(with restrictions in values and types)
and ensures that LLM outputs obey that schema.
Like LangChain, 
the primary mechanism for fixing issues is to
simply retry.
Additionally, there is no guarantee of correctness
for outputs, beyond user-provided data validation rules.
\item LlamaIndex (\url{https://www.llamaindex.ai/}) provides a way to easily inject unstructured information
into prompts to ensure that LLM-context length limitations are met,
targeting data processing applications that leverage a predefined
corpus. Once again, defining the workflow is left up to the user.
\end{tightitemize}
While these solutions offer some conveniences with 
respect to the actual LLM invocations,
defining the actual LLM workflow and ensuring 
accuracy is left up to the user.}

A concurrent preprint explores
the complementary question of whe\-ther LLMs
can be used in the place of the crowd in 
crowdsourcing workflows~\cite{wu2023llms}.
\techreport{Other papers have explored the use of LLMs
for specific unit crowdsourcing tasks~\cite{argyle2023out,gilardi2023chatgpt,tornberg2023chatgpt}, 
while others have found that 
crowd workers often use LLMs~\cite{veselovsky2023artificial}.
Another recent paper confirms that LLMs
do not, in fact, behave exactly 
like crowd workers~\cite{webson2023language},
yet they do show similarities~\cite{dasgupta2022language}.}
Some recent work has recognized
that problem decomposition in LLM workflows
is related to that in crowdsourcing~\cite{wu2022ai},
without actually leveraging
any principles from crowdsourcing.
On the other hand, a recent preprint has applied
techniques analogous to those in crowdsourcing
to determine whether to get additional samples
from LLMs~\cite{aggarwal2023let},
while another has applied
ensembling techniques
across multiple LLMs to reduce cost~\cite{chen2023frugalgpt}.
While these approaches are in the spirit of
what we propose, here, we craft
a broader agenda around leveraging declarative
crowdsourcing principles for prompt engineering.
Finally, a recent paper explores the use
of LLMs for data wrangling~\cite{narayan2022can},
without exploring optimizing the workflow
for cost or accuracy.

\section{LLMs and Crowdsourcing}

\stitle{LLMs and Prompts: A Primer}
This process of leveraging LLMs 
by providing instructions
as part of a textual prompt
is called {\em in-context learning},
and when a few examples are provided, 
it is called 
{\em few-shot learning}~\cite{brown2020language}; 
the alternative is {\em zero-shot}, where no examples are provided.
Standard LLMs take text as input and
provide the next most likely text as output; chat-based LLMs are
a variant that 
are tuned specifically for 
text that looks like a conversation between
a human user and the model. 
The text input can't be arbitrarily long,
and are bounded by an LLM-specific
{\em context length}.
LLM providers include
OpenAI, Cohere, Hugging Face, and Anthropic, 
as well as all major cloud platforms; 
these providers expose an API with 
various parameters in addition to the
text input,
and return a text output. 
Typical parameters include 
the model name and version, temperature 
(i.e., the degree of randomness) and
maximum number of tokens (i.e.,
a constraint on the output length).
\techreport{Other, less frequently used parameters
include those that prioritize the presence
or absence of certain types of tokens
(e.g., those that have appeared before,
or those that are common),
or those that additionally accept 
function definitions---if the output
has to match certain output constraints.
For our purpose, we focus on the most common parameters.}
Some models, while returning a text output,
also return other alternative predicted outputs,
as well as log probabilities 
for each token generated~\cite{logprobs}. 
Pricing for LLMs is based on a per-token rate, 
usually with a different rate
for inputs and outputs. 

\stitle{Shared Terminology} 
To maintain parallels between prompting and crowdsourcing,
we discuss terminologies together.
We are given a {\em problem} on
one or more {\em datasets},
e.g., sorting a collection of text snippets on sentiment.
Each dataset contains multiple {\em data items} or {\em records}.
We refer to a single invocation of the LLM
or the crowd as a {\em task} or {\em prompt},
with the structure of the prompt or task
referred to as the {\em template}:
usually, a template contains some instructions
and/or examples, followed by 
the specific data items to be operated
on by the LLM or crowd worker.
\techreport{Conventionally, in crowdsourcing,
the examples are fixed. 
In LLMs, the examples
could be picked to be similar
to the data items at hand~\cite{liu2021makes,rubin:1987},
or diverse~\cite{su2022selective}.}
Given a task or prompt, the LLM or 
crowd worker returns a {\em response}.
While this response can contain 
other text, such as
an explanation or formatting,
we can automatically extract the underlying
answer to the task.

\stitle{How Crowds are Similar to LLMs}
Since LLMs are trained on massive volumes
of textual data authored by humans,
they can be viewed to be an amalgamation 
of many human opinions~\cite{santurkar2023whose},
and typically complete text in ways that are analogous to human authors~\cite{andreas2022language}. 
Moreover, instruction-tuned LLMs,
which have widely been regarded
as highly successful,
leverage crowdsourced preference data as a key ingredient~\cite{ouyang2022training}.
Asking LLMs
to explain answers rather than simply stating them 
helps them avoid mistakes
in a so-called {\em chain of thought} approach~\cite{wei2022chain,kojima2022large}; similarly
the crowdsourcing literature recommends
using a free text explanation field that the crowd worker needs to complete in order to ensure
they are spending time on the task and not answering randomly~\cite{turk}. 
Prior work in crowdsourcing has shown that
the instructions in the task impacts accuracy,
as does examples~\cite{DBLP:conf/kdd/JainP09,khanna2010evaluating},
just like with LLMs~\cite{Saravia_Prompt_Engineering_Guide_2022}.

\stitle{How Crowds are Different from LLMs}
There are several ways crowd workers are different
from LLMs.
First, there are different knobs for controlling LLMs
and crowdsourcing.
For example, there is no analog
of temperature in controlling the non-determinacy
of outputs with crowds.
With crowdsourcing, one can enforce strict 
output constraints as part of the interface, e.g.,
using drop-downs to restrict
the space of values.
Enforcing constraints 
with popular LLMs is much trickier\techreport{---one 
must do some form of rejection 
sampling to draw observations 
from a constrained space, 
or ask for a JSON output, 
which LLMs are good at understanding, 
but this might shifts the 
calibration of the LLM on the task}.
Second, \techreport{crowds and LLMs respond differently
to instructions.
For example, }small changes in instructions~\cite{zhao2021calibrate}
and examples~\cite{lu2021fantastically}
can have an outsized effect for LLMs,
unlike crowd workers.
Third, crowds and LLMs behave differently.
LLMs have trouble with logical reasoning,
or even basic arithmetic.
They are also prone to {\em hallucination},
where they conjure up new facts 
that are amalgams of existing facts\techreport{,
while crowds are less likely to do so}. 
LLMs also are more likely to accumulate errors 
over the course of generating text~\cite{holtzman2019curious}.
Instead, crowd workers are more prone to 
satisficing, responding 
quickly without reading through the
entire set of instructions,
so that they can get paid 
and move onto the next task.
Crowd workers also make more mistakes
on tasks that require them 
to remember and recall more data,
unlike LLMs that have a higher limit
for this in the form of context length,
even if they do ignore large fractions
of long prompts~\cite{liu2023lost}. 
Third, \techreport{cost models for crowds and LLMs are different,
where }crowds are paid per-task---and varying this price tends to impact the accuracy 
by attracting different sets of crowd workers---while LLMs are charged per-token.

Overall, while there are certainly 
key differences between the two,
there are enough
similarities to justify
leveraging principles drawn
from crowdsourcing to make
LLM-centric data processing workflows 
better.

\section{Declarative Prompt Engineering}
\label{sec:declarative}

We envision a wrapper around a tool like 
LangChain, with individual 
data processing primitives such
as sort, filter, join, categorize, cluster,
find, fill, resolve, ...
to be applied to one or more datasets.
Users can also provide an overall budget
and a desired accuracy.
In addition, users can also provide
``gold standard'' 
test answers
as well as 
special prompt templates that pertain
to the task at hand---these templates 
can also be drawn from repositories or guides~\cite{bach2022promptsource, ding2021openprompt,Saravia_Prompt_Engineering_Guide_2022, openai-cookbook}

Next, we describe a series of principles
drawn from the crowdsourcing literature
that we can leverage and extend for prompt engineering. 
We also include some case studies to illustrate 
benefits of using such principles, leveraging a variety of LLMs, 
problems, and datasets. 
For all case studies, we set the LLM temperature to 0.

\subsection{Varying Prompting Strategies}\label{sec:various-prompting-strategies}
Users of LLMs
typically make use of a single task to accomplish
their entire objective.
For example, if they wanted
to sort a number of textual snippets
on sentiment, they would place them all in one task,
and ask the LLM to rank them.
The crowdsourcing literature
tells us that there are multiple ways
to accomplish the same goal, often
with different cost/accuracy trade-offs.
For the specific task of sorting~\cite{markus-sorts-joins},
instead of having a single task
for sorting the entire set of data items,
one could use pairwise comparisons,
where a given task involves comparing
a pair of data items at a time, with $O(n^2)$
tasks in total.
Another approach involves soliciting
a rating per item, with $O(n)$ tasks, 
and using those
ratings to sort.
In such settings, we would expect
the more fine-grained approach
to perform better than the more coarse-grained one
in terms of accuracy, but cost more.
Similarly, for counting~\cite{DBLP:journals/pvldb/0002KMMO12},
one option is to use coarse-grained tasks
that estimate the proportion
of data items that satisfy some property
(via ``eyeballing''), 
versus those 
that individually check each data item
in a fine-grained manner.
Finally, for entity resolution~\cite{crowder},
one could use a coarse-grained task
that involves grouping multiple data items
into identical sets,
or one could use a fine-grained task 
that involves comparing pairs of data items
at a time.
Therefore, for all these tasks,
if we wanted to guarantee higher accuracy
and cost didn't matter as much,
rather than simply asking the LLM
to resolve, sort,
or count the items, all in one single
task, it would be advantageous to employ
finer-grained unit tasks
that are then aggregated together to give
the answer.

\begin{table}[]
    \centering
    \small
    \begin{tabular}{p{3cm}|p{1cm}|p{1.7cm}|p{1.7cm}}
       {\bf Method}  & {\bf Kendall Tau-$\beta$}  & {\bf \# Prompt Tokens} & {\bf \# Completion Tokens} \\
       \midrule
       Sorting in one prompt &  0.526 & {\bf 152} & {\bf 117} \\
       Coarse-grained ratings & 0.547 & 1615 & 900 \\
       Fine-grained comparisons & {\bf 0.737} & 12065 & 10884 \\
    \end{tabular}
    \caption{Results for sorting 20 flavors, demonstrating the tradeoff between cost and accuracy in prompting strategies. }
    \label{tab:chocolatesort}
    \vspace{-25pt}
\end{table}

\stitle{Case Study on Sorting via Three Prompting Strategies}
Here, we explore sorting
with the three aforementioned approaches:
{\em (i)} listing all the items in the prompt and asking
the LLM to sort the list---our baseline approach,
{\em (ii)} employing $O(n^2)$ pairwise
comparisons each as a separate task, followed by sorting
based on the total number of pairwise comparisons
a given data item ``won'', with ties broken arbitrarily,
and finally {\em (iii)} getting a rating from 1 (least)--7 (most) for each data item each
as a separate task,
and then sorting based on those ratings,
with ties broken arbitrarily.
Specifically, we used the gpt-3.5-turbo model 
from OpenAI to rank 20 ice cream flavors 
by how ``chocolatey'' they are, 
comparing the results 
to a human-labeled ground truth ordering,
verified by two of the authors of the paper. 
The ground truth contained flavors 
with ``chocolate'' in the beginning 
of the list and fruit-related ice creams such as  
``lemon sorbet'' 
at the end. 
For our baseline approach,
we found that the LLM
ordered the flavors with ``chocolate'' in the title at the beginning of the list, while 
the rest of the ordering was seemingly random---this approach had a Kendall Tau-$\beta$ score---a standard metric to compare rankings---of 
0.526.
The pairwise comparison strategy 
required $20 \choose 2$ calls to the LLM, 
and had a score of 0.736,
while the rating-based strategy had
a score of 0.547.
We list the scores and the total input and output
sizes in tokens---which dictate the cost---in Table~\ref{tab:chocolatesort}. 
Overall, we find that different strategies
offer different cost-accuracy characteristics,
with the pairwise approach leading to highest cost
and accuracy, the baseline approach
having the lowest cost and accuracy,
and the rating approach being in the middle
on both fronts.
\takeaway{Rather than trying to accomplish the
entire objective via a single task, 
it is beneficial to explore
other task types, especially
to maximize accuracy.}

\subsection{Hybrid Coarse $\rightarrow$ Fine-grained Prompting}
In our previous case study,
accomplishing the entire task via
one LLM call was viable---even if
it led to a low accuracy overall.
This was because we had a relatively
small number of data items for the LLM
to sort. 
As we increase the number of data items,
it becomes difficult for the LLM
to even complete the task,
despite the fact that LLMs have
high context windows.
The responses are 
error-prone,
with random hallucinations (new data items
introduced) and low recall (old data items
omitted). Recent work has also
shown that LLMs largely ignore
text ``in the middle'' of
long prompts~\cite{liu2023lost}.
In such cases, it is advantageous
to use multiple types of tasks
to accomplish the goal,
often with coarse-grained tasks first,
followed by fine-grained tasks.

Similar strategies have
been applied in crowdsourcing as well.
For example, for finding the max in a
set of items,
Khan et al.~\cite{khan2017dynamic}
employ two kinds of tasks:
a pairwise comparison task
and a rating task.
The latter is more appropriate
at the early stages
for coarse-grained bucketization 
into ratings,
while the former can be employed
for fine-grained comparisons 
amongst those that have higher ratings.
This approach was shown
to have higher accuracy
and lower cost than approaches
that leveraged ratings alone,
and lower cost than approaches that
leveraged pairwise comparisons alone.
Similar hybrid approaches
were applied for sorting~\cite{markus-sorts-joins} as well.
For clustering, Jain et al.~\cite{jain2016s},
employ a two-stage process
of identifying appropriate clustering schemes,
and then categorizing the remaining items
in the clusters.
\longtechreport{Similar
ideas have been applied to search~\cite{datasift}
and taxonomy creation~\cite{DBLP:conf/chi/ChiltonLEWL13}.}


\begin{table}[]
\small
\begin{tabular}{l|l|l|l|l}
{\bf Trial} & {\bf Method}                                    & {\bf Score} & {\bf \# Missing} & {\bf \# Hallucinated} \\
        \midrule

1     & Sorting in one prompt                     & 0.966       & 4                & 1                     \\
1     & Sort then insert  & {\bf 0.999}       & {\bf 0}                & {\bf 0}                     \\
2     & Sorting in one prompt                     & 0.889       & 7                & 0                     \\
2     & Sort then insert & {\bf 0.980}       & {\bf 0}                & {\bf 0}                     \\
3     & Sorting in one prompt                     & 0.940       & 4                & 1                     \\
3     & Sort then insert  & {\bf 0.992}       & {\bf 0}                & {\bf 0}                    
\end{tabular}
    \caption{Results for sorting 100 words in alphabetical order, over 3 trials. Asking the LLM to perform comparisons  to insert words missed in the original sort improves performance.}
    \label{tab:alphabetsort}
    \vspace{-25pt}
\end{table}


\stitle{Case Study on Sorting Many Items: Sort $\rightarrow$ Insert} 
We consider sorting once again,
with a larger number of items,
to investigate if it leads
to hallucinations and
dropped words. 
To allow us to programmatically
consider sorting a large number of items,
we generated 
a list of $n=100$ 
random English words from a dictionary,
with the goal of having the words sorted in
alphabetical order---a task
we would expect the LLM to do well.
We then used the same baseline
approach as in Section~\ref{sec:various-prompting-strategies}, where
we provided
the entire list of words to the LLM 
(Anthropic's Claude 2),
and asked it 
to return the words in sorted order.
Across three trials,
sorting
in a single prompt led to 
between 4--7 words that were missing,
and 0--1 words that were hallucinations
in the LLM response. 
In Table~\ref{tab:alphabetsort},
we also report the Kendall's Tau score
after inserting the missing items in random locations.

To improve on this approach, we then considered
a coarse-to-fine grained hybrid prompting
strategy, where we started
by asking the LLM to sort
the entire list of items.
We then dropped all hallucinated words.
Finally, to reinsert missing words
into the sorted list, 
we asked the LLM to compare 
each missed word to the other words 
in the partially sorted list as pairwise comparisons (i.e., $O\left(kn\right)$ 
calls to the LLM, for $k$ missing words). 
A naive strategy for deriving the index 
where we insert a missed word 
$w_i$ is to find the first sorted word 
that the LLM deems less than $w_i$; 
however, this method can perform poorly 
because the LLM is not guaranteed to 
correctly order each pair of words, 
and a mistake at an early index 
can disproportionately penalize the resulting sort. 
Instead, we compared each word $w_i$ 
to {\em all} sorted words\techreport{,
twice (one prompt with $w_i$ listed first, 
and the other prompt with $w_i$ listed second),
to account for any ordering biases present in the LLM}. 
Then, we derived its index 
by maximizing the ``alignment'' of all of $w_i$'s comparisons.
That is, we picked the location for $w_i$ that
had the least number of pairwise comparison results inverted.
Over 3 trials\techreport{, i.e., different lists of 100 words}, this insertion strategy gave a nearly perfect final sort, with an average Kendall Tau-$\beta$ statistic of 0.990;
and trivially, thanks to the insertion, the number of
missing items at the end of this process is 0.
\takeaway{Employing hybrid strategies, with
coarse-grained tasks first, followed by fine-grained
ones, can lead to low cost and high accuracy overall.}

\subsection{Ensuring Internal Consistency}

When a given problem involves issuing a batch of 
interrelated tasks, 
enforcing consistency across the tasks
can be used to improve the accuracy
on each individual task.
Consider entity resolution, for example,
by having the LLM compare pairs of entities
at a time.
If an LLM says that 
entity A is the same as entity B,
and that entity B is the same as entity C,
then either C is the same as A,
or one of the first two comparisons are incorrect.
Said another way, entity resolution 
with a batch of pairwise entity resolution
tasks must respect transitivity.
This approach of 
ensuring consistency for a batch of tasks
has been applied in crowdsourcing to a few
problems. 
For example, for entity resolution
on bipartite graphs,
which can be formulated as a fuzzy join,
Wang et al.~\cite{wang2013leveraging} leveraged
transitivity to sequence comparisons
in a manner that reduces cost by avoiding
the obvious matches and non-matches.
In a similar vein, one can ensure internal
consistency in sorting and max determination 
problems, e.g.,~\cite{so-who-won},
where, given a set of pairwise comparisons 
(i.e., is a $>$ b), the goal is to identify 
those that are incorrect ones, such that we 
can ensure a global topological sort order 
or global consensus on the max.
Under certain 
accuracy models,
flipping the minimum number of edges
that leads to a topological sort
leads to the maximum likelihood sort order or max item.
Given LLMs randomly make mistakes, we expect
that they often violate internal consistency---and
therefore patching their results after
the fact can help improve accuracy.

\stitle{Case Study on Entity Resolution
while enforcing Internal Consistency} 
We consider applying this idea
to an entity entity resolution 
task on the DBLP-Google Scholar citations 
dataset~\cite{kopke2010paper}.
Prior work has crafted 
a set of questions from this dataset, 
including train, validation, 
and test sets, 
where each question 
compares a pair of citations~\cite{magellandata}. 
We restricted our experiments 
to the validation set of 5742 pairs. 
As a baseline, for each question, 
we asked the LLM (OpenAI's gpt-3.5-turbo model) whether two citations were duplicates.\techreport{ We used the prompt ``Are Citation A and Citation B the same? Yes or No? Citation A is…. Citation B is… Are Citation A and Citation B the same? Start your response with Yes or No.''} 
The baseline method achieved a F1 score of 0.658, with a high precision (0.952) and low recall (0.503). 
For entity resolution tasks, recalling duplicates is quite important, so we wanted to leverage internal consistency to improve recall.

One simple approach to leverage
internal consistency is to flip LLM ``no'' 
responses to ``yes'' if, by transitivity, 
two citations were duplicates.  
\techreport{For example, the papers titled 
``indexing the positions of continuousl...'' and ``bindexing the positions of continuous...'' were not flagged as duplicates by the baseline, 
but both were duplicates of 
``indexing the positions of continously...'' and 
therefore could have their edge flipped 
to be a duplicate.}
Since the validation set is sparse relative
to the number of entities,
the number of transitive edges is quite small.
Therefore, we augmented the validation set 
with additional comparisons.
We used the 
OpenAI text-embedding-ada-002 
model to create an embedding for each entity and determined neighboring citations 
based on L2 distance in embedding space. 
For each question in our dataset, 
where a question has two citations $A$ and $B$, we 
considered the $k$-nearest 
neighbors of each citation 
(i.e., $O\left(2k\right)$ citations), 
and asked the LLM to compare 
each pair of citations within the set of citations and 
its neighbors (i.e., $2k \choose 2$ pairs). 
If the LLM found some ``path'' from $A$ to $B$, 
where an edge between two citations exists 
iff the LLM deems the citations duplicates, 
then we also marked $A$ and $B$ as duplicates, 
even if there was no edge between $A$ and $B$. 
We experimented with $k=1$ and $k=2$, 
finding an increase in F1 score, as shown in Table~\ref{tab:citations}, 
of over 6\%. 
This simple strategy of flipping ``no'' edges based
on transitivity is highly effective in 
improving the F1 score in this setting,
by increasing the recall while slightly 
reducing the precision.
As future work, to 
improve both precision and recall, one could 
consider flipping both ``yes'' and ``no'' edges based 
on whether there is enough evidence in the opposite 
direction.
\takeaway{Fixing erroneous LLM responses based
on evidence from other responses
can be an effective way to improve accuracy.}

\begin{table}[]
    \centering
    \small
    \begin{tabular}{l | c | c | c}
       {\bf Nearest Neighbors}  & {\bf F1} & {\bf Recall} & {\bf Precision} \\
       \toprule
       0 (Baseline)   & 0.658&0.503&{\bf 0.952}\\
       1&   0.706  & 0.569 & 0.930 \\
       2& {\bf 0.722} & {\bf 0.593} &  0.923\\
    \end{tabular}
    \caption{Results for identifying duplicate citations in a slice of the DBLP-Google Scholar dataset. Enforcing consistency between pairwise comparisons amongst more neighbors increases the F1 score.}
    \label{tab:citations}
\vspace{-15pt}
\end{table}

\subsection{Leveraging LLM and non-LLM Approaches}
A useful way to reduce cost is by avoiding LLM 
invocations entirely, especially 
for those tasks where there are 
cheaper proxies.
For example,
say we could tell 
what the LLM response to a given task
would be with high probability
using a low-cost approach,
such as a cheaper, open-source LLM
or other model. Then,
we can avoid asking the LLM,
and instead save the budget
for other tasks that really require it.
For example, for entity resolution,
Wang et al.~\cite{crowder} use
a hybrid human-machine workflow
for entity resolution,
only crowdsourcing
the comparison of pairs
of entities that exceed 
a certain likelihood threshold as determined by a cheap model.
In this case, the cheaper proxy was determined
upfront. In other cases,
this cheaper proxy can be determined
using the LLM/crowd.
For example,
Marcus et al.~\cite{markus-sorts-joins}
leverage the crowd to extract features
of each entity
and use them as a filter to determine whether
the entities need to be compared for the
purpose of entity resolution.
Similarly, Gokhale et al.~\cite{Gokhale:2014wv}
use the crowd to derive blocking rules
for entity resolution,
and then use the crowd for comparing
entities within each block,
following which they train an ML model 
to avoid having to ask the crowd for every entity
pair.

Similar ideas
are applicable to LLMs.
One could use a low-cost non-LLM 
model built
by a human expert\techreport{---and only
ask the hard cases to the LLM}.
Or, 
given LLMs can synthesize programs, 
one could use the LLM
to write code to train a model given
the specific task\techreport{, e.g., entity resolution
or data imputation}.
In either case, the 
low-cost model can be used by default,
and for the cases where there
is uncertainty (as deemed by model
confidence scores), we can leverage
the LLM.

\begin{table}[!t]
    \centering
    \footnotesize
     \begin{tabular}{p{2.5cm} | c | c | c | c}
        & \multicolumn{2}{c|}{{\bf Accuracy}} & \multicolumn{2}{c}{{\bf \# Tokens}} \\
       \cline{2-5}
      {\bf Strategy}  & {\bf Rest.} & {\bf Buy} & {\bf Rest.} & {\bf Buy} \\
       \midrule 
       Naive $k$-NN & 73.26\% &  67.69\% & 0 & 0\\
       Hybrid (no examples) & 84.88\% & 87.69\% & 2838 {\footnotesize $\left(\downarrow 50\%\right)$} & 1624 {\footnotesize $\left(\downarrow 55\%\right)$} \\
       LLM-only  (no examples) &  59.30\% & 81.54\% & 5676 & 3640 \\
       Hybrid (3 examples) & {\bf 89.53\%} & 87.69\% & 7955 {\footnotesize $\left(\downarrow 50\%\right)$} & 5133 {\footnotesize $\left(\downarrow 55\%\right)$} \\ 
       LLM-only (3 examples) & {\bf 89.53\%} & {\bf 92.31\%} & 15910 & 11505 \\
    \end{tabular}
    \caption{Results for a missing value imputation task with a mix of LLM and non-LLM ($k$-NN) strategies. $k=3$ for the $k$-NN algorithm. Even when including $k=3$ neighboring examples in the prompt, boosting LLM performance, the hybrid method achieves similar performance to the ``LLM-only'' strategy, while significantly reducing the number of calls to the LLM. }
    \label{tab:mvimputewithk}
    \vspace{-25pt}
\end{table}

\stitle{Case Study on Data Imputation
by Combining LLM and non-LLM Approaches} 
We explore using $k$-nearest
neighbor ($k$-NN) as a non-LLM strategy,
and focus on data imputation for 
records with a missing attribute value.
Specifically, for each such record,
$k$-NN imputes the missing 
attribute from the 
mode,  or most commonly occurring value,  
of the neighbors' attribute values.
The LLM-based strategy 
 asks the LLM (here, 
Anthropic's Claude model) 
to predict the value 
of the missing attribute 
given a serialized representation 
of the known attributes in the entity. 
\techreport{A serialized entity $e$ with $j$ attributes $a_1\hdots a_j$, 
and values $e_1\hdots e_j$ where $e_j$ is missing is listed as:
$
a_1 \text{ is } e_1; a_2 \text{ is } e_2; \hdots a_{j-1} \text{ is } e_{j-1}.
$}
We finally considered
a hybrid approach where 
we use the value imputed by $k$-NN 
if all neighbors contain the same value 
for the missing attribute---otherwise, 
the LLM is prompted to return the missing value.
Since, in this case, we have a set of 
records with have ground truth values (used by $k$-NN),
we could optionally insert examples
into the prompt for the LLM,
in both the Hybrid and the LLM-Only approaches,
since examples can help improve accuracy~\cite{liupretrain2023},
while also increasing cost (in terms of input tokens).
Overall, we consider five approaches: $k$-NN,
LLM-Only (with no examples), Hybrid (with no examples
on LLM calls), LLM-Only (with $k'=3$ examples),
and Hybrid (with $k'$ examples
on LLM calls),
with the results shown in 
Table~\ref{tab:mvimputewithk}
on the Restaurants and Buy datasets~\cite{Mei2021CapturingSF}.

Focusing on the no-examples setting first, we find
that the hybrid method
outperformed both the non-LLM and LLM-only strategies,
at roughly half the cost of the LLM-only strategy.
Notably, the hybrid approach performed
better than the LLM-only approach
in this case, because the LLM-only approach
sometimes imputed values
that did not match exactly
with 
the ground truth (e.g., ``TomTom'' instead of ``Tom Tom'' or ``Elgato Systems''  instead of ``Elgato''),
and thus may have been unfairly penalized.
When examples are added, 
both the hybrid and LLM-only 
strategy get more expensive,
but the hybrid approach manages
to achieve an identical performance on one
dataset, and a slightly worse performance on the other,
at roughly half the cost of the LLM-only strategy.
\takeaway{Leveraging a non-LLM proxy can help substantially
reduce costs while keeping accuracy similar.}

\techreport{
\subsection{Quality Control}
}

\techreport{
The main approach for ensuring
accuracy with LLMs
is to check if the LLM
outputs an answer
that violates certain syntactic
constraints (e.g., 0 for a Yes/No answer),
and then retry the query.
We propose drawing on ideas from crowdsourcing
to make this more principled.
The first challenge is
even understanding what the
accuracy of
the LLM is for a given type of task.
Following best practices in the crowdsourcing
literature~\cite{turk,chen2011opportunities}, one to do this is 
via a validation set,
for which the ground truth answer is known---and
one can infer the accuracy based on 
the fraction correct.
In the cases where a validation
set is unavailable, we can apply
expectation-maximization type approaches~\cite{quality}
across a set of LLMs for the same task,
where the underlying assumption is that each
LLM answers the task independently,
and has a fixed, but unknown accuracy
for that type of task.
Other approaches to quality control
include verification~\cite{little2010turkit,soylent}
the LLM verify its own response
as a followup, or have another LLM
do so. Similar ideas
have demonstrated promise in LLMs~\cite{madaan2023self}.
Finally, one can also try to
debias or better calibrate LLM answers, just like
one would in crowdsourcing~\cite{rion-snow,DBLP:conf/kdd/ZhuangPRH15}---there have been some early attempts
at calibrating LLMs~\cite{zhao2021calibrate},
but more remains to be done.
}

\techreport{
Once we know
about the accuracy
for each LLM
for a given task,
we can then apply
techniques that determine
which LLM to ask
at each step,
to ensure
a given accuracy overall,
while keeping costs low.
Of particular note are 
probabilistic approaches
drawn from crowdsourcing~\cite{crowdscreen},
where we determine,
based on the answers so far,
whether it is worth
asking another LLM,
or to finalize an answer---data items
for which there is more disagreement
across LLMs
or less confidence from each LLM
are more valuable to spend
money on rather than 
those for which there is high confidence
or agreement across LLMs.
Similar approaches have
demonstrated promise
for reasoning tasks, even for a single LLM~\cite{wang2022self},
where multiple reasoning paths
are extracted, following by
a majority vote
to arrive at the final answer.}

\section{Discussion}

While we can 
draw from the crowdsourcing 
literature when thinking about 
how to reliably decompose 
LLM-native workflows into subtasks, 
as in crowdsourcing, 
there is no ``one-size-fits-all'' strategy. 
In this section, we discuss 
some practical considerations 
necessary to achieve our research vision 
of a declarative prompt engineering toolkit 
for a variety of data-related tasks, 
as well as complementary work 
from ML communities around prompting strategies.

\stitle{Identifying Best Prompting Strategies Automatically} 
It's relatively easy to decompose 
data-related workflows into primitives such as ``sort'' and 
``filter'', but there are many 
algorithms to perform these primitives, 
all with various tradeoffs between cost and accuracy, as 
described in Section~\ref{sec:declarative}. 
There are several lingering questions, such as: which 
algorithms get us ``good-enough'' performance?
Do we need to improve the performance 
by ensuring internal 
consistency? 
For processing a subset of records, 
can we rely on non-LLM strategies to save cost? 
Answering these questions 
requires experimentation on a 
sample of the dataset, 
much akin to the train-val-test split 
paradigm in traditional machine 
learning~\cite{shankar2022operationalizing}. 
We envision users of a declarative prompt
engineering
toolkit to label a small number of 
records as a ``validation set,'' 
which can be used to explore the 
cost-accuracy tradeoff for the user's specific workflow. 
Similar to AutoML, a declarative prompt
engineering toolkit can shoulder 
the burden of evaluating all 
strategies and recommend a strategy to apply to the entire 
dataset, given a user-defined budget.  
Additionally, the toolkit can consider hyperparameters, such as batch size, as other dimensions 
to optimize over: for example, 
one can ask the LLM to process 
a small number of comparison tasks 
in a single prompt, reducing cost and latency 
with implication on accuracy.
We could also leverage prompt repositories~\cite{ding2021openprompt,bach2022promptsource}
to identify other prompting strategies
and templates
and explore/evaluate them automatically.
Recent work has also demonstrated 
that one can explore a space of prompts automatically to pick the best ones from a quality standpoint~\cite{shin2020autoprompt,zhou2022large}.

\stitle{Mitigating Prompt Brittleness} 
Prior work has noted the brittleness of prompts, 
where slight changes 
in wording drastically alters overall 
accuracy~\cite{zamfirescu2023johnny}. 
Moreover, the effectiveness of a given prompt
can also vary greatly between models~\cite{langchain-problem}. 
While prompts may prove to be less brittle 
in the long run, as LLMs are trained on a larger variety of 
prompts, using LLMs for data-related tasks 
{\em now} requires us to consider best practices for reliably 
extracting information from an LLM response.
The applied LLM community has put forth primers on techniques 
such as chain-of-thought prompting, 
where instructing the LLM to 
``think step by step'' in the prompt 
produces longer and more 
accurate results~\cite{wei2022chain}. 
Of course, this is at the 
expense of additional tokens, 
pointing back to exploration of 
the cost-accuracy tradeoff that a 
declarative prompt engineering toolkit 
should help its users understand. 
Moreover, chain-of-thought prompting 
and similar techniques that draw out LLM responses with 
no particular structure make it 
difficult to transform an LLM's 
response into the answer programmatically. 
For example, asking a multiple choice question and instructing the LLM to explain why they chose an 
answer might yield a response
including instances of each answer 
in the text, e.g., ``I choose A because B and C are not relevant, and D is not accurate.'' 
If not every answer begins with ``I choose,'' and some LLMs might put the chosen answer at the end of the response, which answer choice do we extract? 
In fact, for our entity resolution case study,
we found that experimenting with chain-of-thought would
sometimes lead to the LLM outputting: "They are not the same...[explanation]...They are the same." 
Additional subtasks to infer the answer 
from another LLM's response might be necessary, 
if LLM response structure is not 
well-defined or consistent among multiple calls,
but this will add to the cost.
More recently, LLMs have been shown to reliably output information in specific data formats, such as JSON, when explicitly prompted to~\cite{white2023prompt}. 
Extracting structured data allows us to apply well-known data management for ML techniques, such as validating schemas and constraints on specific attributes~\cite{breck2019data}, to outputs of an LLM, improving their reliability and thus our confidence in using them.

\section{Conclusion}
We presented a new research agenda
around making it easier
to optimize
LLM-powered data processing workflows
for cost and accuracy,
when run at scale.
The declarative crowdsourcing
literature provides a rich set of techniques
for handling cost and accuracy tradeoffs
when dealing with noisy oracles---and
all of these techniques could,
at least in theory,
be applied to LLM-powered workflows as well.
We illustrate via case studies
that some of these techniques
do provide benefits
in either accuracy or cost or both.
We therefore believe the database community
has a key role to play in making 
the vision of declarative prompt engineering a reality.

\techreport{
\stitle{Acknowledgments}  We acknowledge support from grants DGE-2243822, IIS-2129008, IIS-1940759, and IIS-1940757 awarded by the National Science Foundation, an NDSEG Fellowship, funds from the Alfred P. Sloan Foundation, as well as EPIC lab sponsors: G-Research, Adobe, Microsoft, Google, and Sigma Computing. The content is solely the responsibility of the authors and does not necessarily represent the official views of the funding agencies and organizations.}

\bibliographystyle{abbrv}
\bibliography{ref}

\end{document}